# ILC ENERGY UPGRADE PATHS TO 3 TEV*

H. Padamsee, Cornell University, Ithaca, NY, 14850, U.S.A

*Abstract*

We consider several ILC energy upgrade paths beyond 1 TeV depending on the needs of high energy physics. Parameters for four scenarios will be presented and challenges discussed.

1. **From 1 TeV to 2 TeV based on:**
A. Gradient advances of Nb cavities to 55 MV/m anticipated from on-going SRF R&D on Nb structures.
B. Radically new travelling wave (TW) superconducting structures optimized for effective gradients of 70+ MV/m, along with 100% increase in *R/Q* (discussed in more detail in paper **WEOCAV04 at this conference**. The large gain in *R/Q* has a major beneficial impact on the refrigerator heat load, the RF power, and the AC operating power.

OR

2. **From 1 TeV to 3 TeV based on:**
A. Radically new travelling wave (TW) superconducting structures optimized for effective gradients of 70+ MV/m, along with 100% increase in *R/Q*. The large gain in *R/Q* has a major beneficial impact on heat load, RF power, and the AC operating power.
B. 80 MV/m gradient potential for Nb3Sn with *Q* of $1 \times 10^{10}$, based on extrapolations from high power pulsed measurements on single cell Nb3Sn cavities. Further, the operating temperature is 4.2 K instead of 2K due to the high $T_c$ of Nb3Sn.

## INTRODUCTION

Over the last few years, there has been general agreement [1] in the World High Energy Physics community that an electron-positron collider Higgs factory is one of the highest priorities for the field. In June 2020, the European Strategy for Particle Physics Report [2] offered strong support for ILC hosted by Japan, expressing their wish for European participation. Other paths to the Higgs Factory are the FCCee [3] or CLIC [4] in Europe, and CepC [5] in China, all in CDR stage with further development needed. With a TDR completed some years ago [6], the superconducting ILC remains the most technologically ready and mature of all possible Higgs factories options for an expeditious start. In the years after its TDR completion, ILC technology is being used on a large scale to establish a rich experience base with new accelerators such as European XFEL [7], LCLS-II [8] in the US and SHINE [9] in China, along with SRF infrastructure installed worldwide. The most significant development supporting the expeditious launch of ILC is that the cost of starting at 250 GeV as a Higgs Factory [10, 11] has dropped considerably (40%) from the original TDR estimate for the 500 GeV machine, with bottoms-up cost evaluations, substantiated by the experiences of EXFEL and LCLS-II. At 17.5 GeV, EXFEL is an SRF linac based on ILC technology that has been operating for several years. The average maximum cavity gradients for 400 cavities as received and prepared by the ILC recipe reached 33.3 ± 6.6 MV/m [12, 13]. (The other 400 cavities which were treated with a different recipe reached 29.8 ± 6.6 MV/m.) Demonstration of gradients > 30.5 MV/m in full scale cryomodules at KEK [14] and a CM gradient > 32 MV/m has been achieved at Fermilab [15] with beam.

Demonstrations at ATF2 (KEK) in Japan have established confidence in ILC IP parameters [16]. Demonstrations at CESR (Cornell) have established confidence in damping ring parameters [17].

A strong physics attraction of ILC is the energy upgradability to TeV and multi-TeV energies, offering clean e+e- physics *to the next century.* All energy upgrade paths will require intense SRF R&D to realize the very high gradient and high *Q* performances needed. But there are several decades of R&D ahead to accomplish those goals before the time for a 2 TeV or 3 TeV upgrade is indicated by physics. We are optimistic that the Snowmass [18] process in progress will stimulate funding for these avenues for high energies.

## OPTION 1A: 2 TEV WITH 55 MV/M

We consider advances in SRF performance to gradients/*Q* of 55 MV/m/$2 \times 10^{10}$ based on the best new treatments [19, 20] applied to advanced shape structures such as the Re-entrant [21], Low-Loss/ICHIRO[22,23] or the Low-Surface-Field (LSF) candidates [24] for which gradients of 52 – 55 MV/m with Q > $10^{10}$ have already been demonstrated with 1-cell cavities, using the standard ILC recipe[25]. The new shapes were developed to reduce $H_{pk}/E_{acc}$ 15 – 20% below that of the TESLA shape. In addition, the *R/Q* for the advanced shapes is about 20% higher to help reduce the RF power, dynamic heat load and AC power.

Today the best result for a 1-cell cavity of standard TESLA shape given the best new treatment is 49MV/m [19,20], confirmed by retesting at many labs, and by about 50 tests on many 1-cell cavities [26]. Therefore, applying the best new treatments to the advanced shapes we can optimistically expect gradients 15% higher with successful R&D, so from 60 – 65 MV/m for single cells and 55 MV/m for 9-cells.

The strategy adopted for Option 1a is to replace the lowest gradient (31.5 MV/m) 0.5 TeV section of cavities/cryomodules, re-using the tunnel, RF and Refrigeration of this section, keep the 0.5 TeV section of the 1 TeV upgrade with 45 MV/m gradient (11,000 cavities), running with the slightly lower bunch charge (Table 2), and add 1.5 TeV with 55 MV/m/*Q*= $2 \times 10^{10}$. With this approach, it is possible to keep the total linac length to 52 km - well below the currently expected 65 km site limit [27]. Note: If we just add a full one TeV (24km)

---

to the existing 1 TeV (38km), the total linac length comes too close to 65 km.

Table 1 shows high level parameters for the 2 TeV upgrade as compared to 1 TeV in the ILC TDR. The luminosity is 7.9 x$10^{34}$ which is higher than the 3.75 x$10^{34}$ for CLIC 1.5 TeV [28]. Table 2 gives more detail parameters for beam and accelerator. The number of particles per bunch is slightly lower than for the 1TeV case, but the number of bunches and rep rate are the same. The peak beam current is therefore slightly lower. The total beam power for two beams increases from 27 MW to 47 MW. Other beam parameters are adjusted so that the spot size at collision is reduced to 1.6 nm (from 2.7 nm).

As shown in Table 2, the total number of *new cavities* at 55 MV/m required for 1.5 TeV is 27,000 spanning a linac length of 36 km, of which 22 km can be installed into the empty tunnel (from the removed 0.5 TeV), leaving 14 km of new tunnel to be installed. Adding in the length (16km) of the 0.5 TeV section remaining with 45 MV/m cavities, the total linac length will be 52 km, below the expected site limit of 65 km. There are savings from cryomodule parts if the tear-down and replacement are staged so that some of the removed cryomodules parts are re-used. From 1600 CMs removed from the 0.5 TeV section, we estimate the parts savings to be in the range of 0.5B, provided the removal and production of CMs are properly staged. For the new 1.5 TeV section, the cavity loaded $Q$ is 6.7x$10^6$, the input power per cavity will be 365 kW, with RF pulse length 2.0 ms, similar to the RF pulse length for 1 TeV. The total number of klystrons required is 1150 of which 360 klystrons are re-used from the 0.5 TeV removed section, and 65 klystrons are available from the 0.5 TeV remaining section (which operates with the new, lower bunch charge), leaving 725 new klystrons to be added. We use 65% efficiency for RF systems installed for 1TeV and above, from R&D on improved klystrons, and 50% efficiency for the RF system installed for the first 0.5 TeV, to give an average RF efficiency of 60%. The total 2 K refrigeration required will be 66 kW, of which 33 kW is re-used, leaving 33 kW new refrigeration to be installed. We assume a cryoload safety factor and RF power overhead of 20% each for the new installations. The damping ring and positron source will be same as for 1 TeV, due to the same number of bunches, but the beam dump cost will increase. Summing all the cost components outlined, the additional cost for the 2 TeV upgrade will be 6.0 B. The AC power to operate at 2 TeV will be 345 MW, making ILC with SRF an attractive path for high energies.

## OPTION 1B : 2 TEV WITH 70 MV/M TRAVELLING WAVE NB STRUCTURES

As discussed in another paper at this conference [29], Travelling Wave (TW) structures offer several advantages compared to standing wave (SW) structures: substantially lower peak magnetic ($H_{pk}/E_{acc}$) and lower peak electric field ($E_{pk}/E_{acc}$) ratios, together with substantially higher R/Q (for lower cryogenic losses, lower RF power and lower AC power). Instead of using the TESLA shape for the cells, the Low-Loss shape further reduces the peak surface magnetic field. In addition, it becomes possible to lower the cavity aperture (from 70 mm to 50 mm) without incurring the penalty of higher wakefields since the beam bunch charge for the 2 TeV upgrade is somewhat *lower* than the bunch charge for 0.5 and 1 TeV stages (Table 1), while the luminosity for 2 TeV is still 2x than for CLIC 1.5 TeV, allowing for even lower bunch charge if necessary. By combining these steps, it becomes possible to obtain an overall 48% reduction in $H_{pk}/E_{acc}$ and factor of 2 gain in $R/Q$ over the TESA standing wave structure. Refs [29] discusses the challenges to develop the TW structures. The TW cavity development effort has started. We expect the cost of TW SRF cavities will be 30% higher due to the recirculating Nb waveguide, leading to 15% increase in the cost of CM for TW structures.

The first strategy to consider in the TW option is again to remove the lowest gradient (31.5 MV/m) 0.5 TeV section, re-use the tunnel, RF and Refrigeration of this section, and keep the 0.5 TeV section (11,000 cavities) section installed for the 1 TeV upgrade with 45 MV/m gradient (but running with the slightly lower bunch charge for the 2 TeV case). Then add 1.5 TeV with TW SRF cavities at 70 MV/m/$Q$= 2x$10^{10}$ and $R/Q$ 2x higher than SW Nb cavities. With this approach it is possible to keep the total linac length to 44 km, well below the currently expected 65 km site limit.

As shown in Table 2, the total number of *new TW cavities* at 70 MV/m required is 21,000 spanning a linac length of 28 km, of which 22 km can be installed into the empty tunnel (from the removed 0.5 TeV), and so requiring 6 km of new tunnel to be installed. Adding in the length (16 km) of the 1 TeV section remaining, the total linac length will be 44 km, well below the expected site limit of 65 km. For 1600 CMs removed from the 0.5 TeV section, we estimate the savings in re-used parts to be in the range of 0.5B, provided the removal and production of CMs are properly staged. For the new 1.5 TeV section, the cavity loaded $Q$ is 5x$10^6$, the input power per cavity will be 460 kW, with RF pulse length 1.76 ms. The total number of klystrons required is 1180 of which 360 klystrons are re-used from the 0.5 TeV removed section, and 65 klystrons are available from the 0.5 TeV remaining section (because it operates with the lower bunch charge than for 1 TeV), leaving 755 new klystrons to be added. The average RF power efficiency of new RF systems will be 0.65 and the existing RF systems from the first 0.5 TeV installation will be 0.5, giving an overall RF efficiency of 0.61. The total 2 K refrigeration required will be 37 kW, of which 33 kW is re-used, leaving 4 kW new refrigeration to be installed. We assume a cryoload safety factor and RF power overhead of 20% each for the new installations. The damping ring and positron source will be same as for 1 TeV, due to the same number of bunches, but the beam dump cost will increase. Summing all the cost components outlined, the additional cost for the 2 TeV upgrade will be 4.9 B. The AC power to operate 2 TeV will be 315 MW, making this path attractive due to the improved environmental impact. Note the substantial benefit to the AC power due to the 2 x higher $R/Q$ of the TW cavities.

A better strategy may be to remove the entire 1 TeV linac, keeping the RF, tunnel and Refrigerator, to install a brand new linac using 70 MV/m TW cavities, we will need to populate the existing 38 km of tunnel with 28,000 TW cavities (no new tunnel needed), and use the existing Refrigeration (no new refrigeration needed), adding 755 klystrons. Savings from re-using CM parts from > 3000 CM from the 1 TeV section is estimated to be 1 B. The additional capital cost for this path will be 5.2 B, comparable to the path above, but the <u>AC power will be 240 MW</u>, much less than the path above. The shorter tunnel and lower AC power may dominate the choice of this path, although it requires much more labor to remove and replace the entire installed linac for 1 TeV.

## OPTION 2A : 3 TEV WITH 70 MV/M TW STRUCTURES

The beam bunch charge for the 3 TeV upgrade is chosen to be *3 x lower* than the bunch charge for 0.5 TeV stage to obtain a luminosity comparable to CLIC 3TeV. The lower bunch charge helps with wakefields and with IP backgrounds. The number of bunches per RF pulse is doubled to 4900, and the bunch spacing is lowered due to the lower bunch charge (See Table 2).

The strategy adopted here is to remove ALL the installed cryomodules for 1 TeV and replace them with new 70 MV/m TW cavities/cryomodules, plus add new linac sections to reach 3 TeV energy. Re-use the existing RF and Refrigeration and CM parts from the removed 1 TeV section. As shown in Table 2, a total of 43,000 TW cavities will be required, so that with the (cavity to linac tunnel) filling factor of 0.75, the total length of the 3 TeV linac will be 57 km, under the expected site limit of 65km. 38 km of tunnel is already present from the 1 TeV removed, requiring 19 km of new linac tunnel. The total number of klystrons required will be 1500, of which 820 are available from the 1 TeV installation. The RF system cost will be higher due to the longer RF pulse length. Also, the existing 820 klystrons and RF system will have to be upgraded to provide longer RF pulses, which will incur a cost of about 0.4 B. The efficiency of the first RF system installed with 360 klystrons for 0.5 TeV is 0.5, and for the later installed RF system for the next 0.5 TeV with 460 klystrons is 0.65, due to improved klystrons. Hence the average RF system efficiency used is 0.61. The input power per cavity will be 300 kW due to the high gradient. The loaded $Q$ will be $8 \times 10^6$. The total 2 K refrigeration requirement will be 95 kW of which 51 kW is already present, leaving a balance of 44kW to be installed. Add in the cost of needed damping rings, positron source and beam dump for increasing the number of bunches from 2450 to 4900. The total *additional* capital cost for 3 TeV (from 1 TeV) will be 11.8 B, shown in Table 1. The total AC power to run 3 TeV will be 400 MW, which is much lower than for CLIC 3 TeV (590 MW).

A lower cost alternative is to only remove/replace the cavities/cryomodules in the first 0.5 TeV of the baseline stage which has relatively low performance (31.5 MV/m), as for the 2 TeV case above. The total number of new cavities installed will be 36,000, to require a tunnel length of 48 km plus 16 km of existing 0.5 TeV to make the total tunnel length of 64 km which is too close to the expected site limit. Therefore, this option is not preferred, despite the slightly lower cost.

Table 2 gives detail parameters (for beam and accelerator) for ILC3 TeV (Option 2a) with 70 MV/m TW structures as compared to CLIC 3 TeV. Note that the backgrounds at the IP for the ILC 3 TeV are much lower than for CLIC, and final beamstrahlung energy spread is 16% compared to 35% for CLIC. To reach the desired luminosity, the beam power is 61 MW with twice the number of bunches (4900) spaced closer together in the linac (250ns instead of 366 for 1TeV) as allowed by the lower bunch charge. The peak beam current is 4.16 mA. The final vertical spot size is 1 nm comparable to the CLIC case.

Figure 1 shows the rough breakdowns for the costs of various systems: Cryomodules, RF, Refrigeration, Conventional Facilities, Damping Rings and Positron Sources for two of the upgrade paths discussed here (1b and 2a) from 1 TeV (TDR) to 2 TeV and from 1 TeV to 3 TeV.

## OPTION 2B : 3 TEV WITH 80 MV/M NB3SN STRUCTURES AT 4.2 K

Option 2b for 3 TeV is to consider 80 MV/m Standing Wave Nb3Sn TESLA-like structures at 4.2 K (due to higher $T_c$ of the new material) with $Q$ values of $1 \times 10^{10}$. In this case the challenge is to develop high performance Nb3Sn which has a much higher (nearly 2X) fundamental critical magnetic field than Nb. Major breakthroughs in materials science and technology will be required, as best gradients today are only at the 20 MV/m level. Due to the combined improvement of Carnot and technical efficiency at 4.2 K over 2 K, the ratio: AC power/cryo power improves from 730 to 230. We assume that the capital cost of 4.2 K refrigeration will be a factor 3 lower than for 2 K, and that the refrigerator units installed for 1 TeV are designed so that 1 watt of cooling at 2 K would be later equivalent to 3 watts of cooling at 4.2 K when the conversion is made for the 3 TeV upgrade at 4.2 K.

The strategy here is to remove all the cryomodules for 1 TeV and replace them with new 80 MV/m – $Q = 1 \times 10^{10}$ cavities/cryomodules, plus install new linac sections to reach 3 TeV energy. Re-use the RF, Refrigeration and CM parts of the removed 1 TeV section, converting the 2K refrigeration to remove heat load at 4.2 K. A total of 37,500 Nb3Sn cavities will be required, so that with the filling factor (cavity to tunnel length) of 0.75, the total length of the 3 TeV linac will be 50 km, well under the expected Japan site constraint of 65 km. 38 km of tunnel already installed for 1 TeV is re-used, so that 12 km of new linac will be required. The total number of klystrons required will be 2500, of which 820 are available from the removed 1 TeV installation. The existing klystrons and RF system will have to be upgraded to provide longer RF pulses (2.6 ms), which will incur a cost of about 0.4 B. The number of new klystrons required is 1680. The average efficiency of old and new RF systems will be 0.63. The input power per cavity will be 550 kW, at a loaded $Q$ of $1 \times 10^7$, so couplers will need to be improved, and microphonics stronger. The total 4.2 K refrigeration required will be 352 kW of which 51 kW (at 2 K) is already present for 1 TeV, equivalent to 150 kW at 4.2 K. The balance of 200 kW at 4.2 K needs to be installed. Add in the cost of needed damping rings, positron source and beam dump for increasing the number of bunches from 2450 to 4900. The total additional capital cost for 3 TeV will be 11.0 B, as shown in Table 1. The total AC power to run 3 TeV will 525 MW.

Incidentally, if the alternative path considered is to install Nb3Sn cavities for 2.5 TeV, leaving 16 km of the 0.5 TeV linac with 45 MV/m gradient in place, the total number of new cavities installed will be 31,000, to require a tunnel length of 41 km. Of this, 22 km is available and 19 km will be new tunnel. Therefore the total linac length will become 16 + 41 = 57 km, quite close to the expected site limit of 65 km, making this path not preferred - despite the 0.5 B cost savings due to fewer cavities.

## AC POWER DEMANDS FOR 2 TEV AND 3 TEV

ILC Energy upgrades beyond 1 TeV require 300 – 400 MW AC power for operation (except path 2b), which reflects the major advantage of the SRF technology. We can expect further reductions in AC power from on-going developments under Green-ILC [30] paramount in importance. Efforts under this umbrella are preparing to explore multiple paths to make ILC and its upgrades environmentally sustainable. Wind power is one avenue following the example of ESS in Sweden [31]. A 30 – 40 units wind turbine farm is capable of providing 100 MW at a cost of 150MEuro. Combined heat and power production using bioenergy or solar photovoltaic cells integrated in the buildings are other examples. New ways of recycling low heat water (below 50°C) would also enable agricultural use of recycled heat, such as greenhouse heating.

## ANTICIPATED COST REDUCTIONS APPLIED

The 1 TeV upgrade discussion in the ILC TDR does not apply any learning curve cost reduction to cavity, cryomodules or klystrons. Between the baseline ILC at 0.25 TeV and the upgrade options to 2 TeV and 3 TeV the total number of cavities increases by a factor of 5 from 8000 to about 40,000, and the total number of klystrons increases by a factor of 5.6 from 250 to 1500. Accordingly, we have applied here a 25% cost reduction for cavities and klystrons for 2.5 doublings, using the 90% learning curve in the TDR. We further assume that due to RF power developments, the efficiency of klystrons will improve from 65% (TDR) to 85%. Taking into account modulator and distribution efficiencies of 90% each, we use 65% efficiency for newly installed RF systems for 1 TeV, 2 TeV and 3 TeV upgrades, but continue to use 50% efficiency for RF systems installed for the first 0.5 TeV. We expect further cost reductions from several areas of R&D already started. Among the areas under exploration are niobium material cost reduction (15 - 25%) for sheet production directly from ingots (large grains), and/or from seamless cavity manufacturing from tubes with hydroforming or spinning to reduce the number of electron beam welds and weld preparations (15 - 20%). Based on the above ideas, we use an overall cost reduction of 50% in the cost of large productions of SW cavities. After including these reductions, we expect the cost of TW cavities will be 30% higher, leading to 15% increase in the cost of CM for TW structures.

Cost-reducing features for cryomodules [32] are to connect cryomodules in continuous, long strings similar to cryostats for long strings of superconducting magnets, saving the cost for the expensive ends. The elimination of the external cryogenic transfer line by placing all cryogenic supply and return services in the cryomodule also reduce costs, not only directly for the cryogenic components, but also by reducing tunnel space required. We estimate that by this method the filling factor from cavities to "linac tunnel length" will improve from 0.7 to 0.75.

## CONCLUSION

Anticipated advances in SRF performance to 70 - 80 MV/m will enable the ILC and its energy upgrades to offer a rich, varied and flexible physics program to complement that of the HL-LHC, and possibly open fundamentally new insights beyond the capabilities of the HL-LHC. The high luminosity with polarized electron beam and low backgrounds give access to rare processes. The clean experimental environment and absence of triggers in high-energy e+e collisions and the good knowledge of the initial state allow precise measurements. New physics has been unsatisfactorily absent from LHC so far, and so precision physics from a lepton collider becomes key to Beyond Standard Model physics. The flexibility and large accessible energy range, almost one order of magnitude, provides a wide range of possibilities for new physics to the next century.

## ACKNOWLEDGMENTS

The author wishes to express many thanks to Sam Posen, Sergey Belomestnykh, Shin Michizono, Akira Yamamota, and Kaoru Yokoya for their comments, suggestions, support and help.

Table I. High level parameters for ILC energy upgrades. Costs do not include Detector and Manpower.

|  |  | ILC1 From TDR | ILC2 Option 1a | ILC2 Option 1b TW | ILC3 Option 2a TW | ILC3 Option 2b Nb3Sn | CLIC 3 |
|---|---|---|---|---|---|---|---|
| Energy | TeV | 1 | 2.0 | 2.0 | 3.0 | 3.0 | 3.0 |
| Luminosity | X10³⁴ | 4.9 | 7.9 | 7.9 | 6.1 | 6.1 | 5.9 |
| AC Power | MW | <300 | 345 | 245 -315 | 400 | 525 | 590 |
| Cap Cost (Total) | BILCU | +5.5 (13.3) | +6.0 (19.3) | +4.9 – 5.2 (18.2-18.5) | + 11.8 (25.1) | + 11.0 (24.3) | 24.2 BCHF |
| Gradient new linac | MV/m | 45 | 55 | 70 | 70 | 80 | 72/100 |
| Q of new linac | 10¹⁰ | 2 | 2 | 2 | 2 | 2 (at 4.2K) | 5700 |
| Av. CM unit cost | M$ | 1.85 | 1.15 | 1.32 | 1.32 | 1.15 |  |

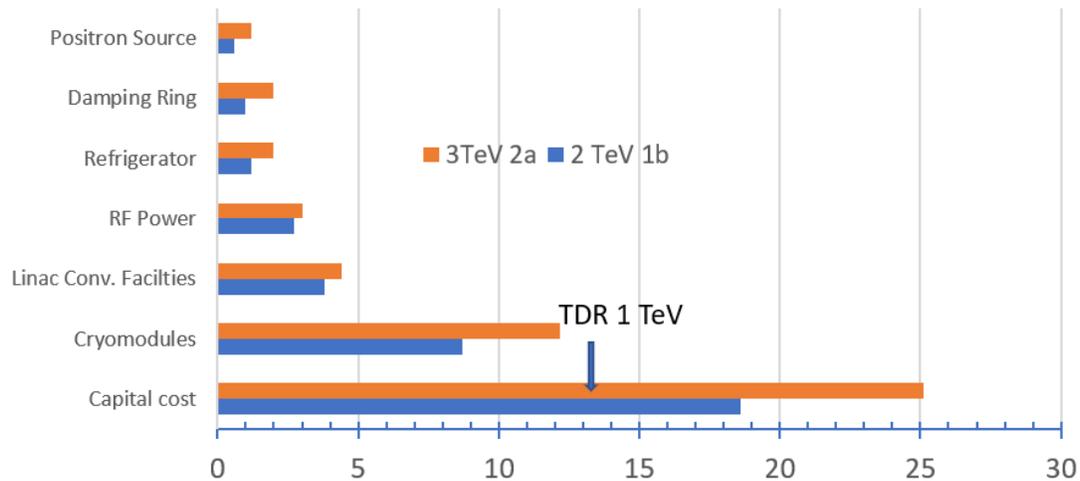

Figure 1: Cost breakdowns for some of the major systems for ILC 2 TeV (Option 1b) and 3 TeV (Option 2a) upgrades beyond 1 TeV. The bars show the TOTAL costs for (1 TeV +2 TeV) OR (1 TeV + 3 TeV). The added costs over 1 TeV are 4.9 B and 11.8 B. The ILC TDR estimates the capital cost for 0.5 TeV as 7.8BILCU and the added cost for upgrading from 500 GeV to 1000 GeV to be 5.5 BILCU.

Table II. Detail Parameter Sets for proposed ILC 2 TeV and 3 TeV energy upgrades from 1 TeV, compared with CLIC 3 TeV.

| | | ILC1 | ILC2 Option 1a | ILC2 Option 1b TW | ILC3 Option 2a TW | ILC3 Option 2b Nb3Sn | CLIC 3 |
|---|---|---|---|---|---|---|---|
| Energy | TeV | 1.0 | 2.0 | 2.0 | 3.0 | 3.0 | 3.0 |
| No. of particles/bunch | $\times 10^{10}$ | 1.74 | 1.5 | 1.5 | 0.65 | 0.65 | 0.37 |
| No. of bunches | | 2450 | 2450 | 2450 | 4900 | 4900 | 312 |
| Bunch spacing | $\times 10^{-9}$ s | 366 | 366 | 366 | 250 | 150 | 0.5 |
| Pulse current | mA | 7.6 | 6.6 | 6.6 | 4.16 | 4.16 | |
| Rep Rate | Hz | 4 | 4 | 4 | 4 | 4 | 50 |
| RF pulse length for added linac | ms | 1.94 | 2.0 | 1.76 | 2.6 | 2.6 | 0.00024 |
| Beam Power 2 beams | MW | 27.2 | 47 | 47 | 61 | 61 | 28 |
| $\varepsilon_x/\varepsilon_y$ (m) | $\times 10^{-8}$ | 500/3 | 500/2 | 500/2 | 500/2 | 500/2 | 66/2 |
| $\beta_x/\beta_y$ (m) | $\times 10^{-3}$ | 22/0.23 | 22/0.23 | 22/0.23 | 16/0.15 | 16/0.15 | |
| $\sigma_x/\sigma_y$ (m) | $\times 10^{-9}$ | 335/2.7 | 237/1.6 | 237/1.6 | 165/1.0 | 165/1.0 | 40/1 |
| $\sigma_z$ (m) | $\times 10^{-3}$ | 0.225 | 0.225 | 0.225 | 0.1 | 0.1 | 0.044 |
| $\Psi$ (Beamstr. Par.) | | 0.21 | 0.5 | 0.5 | 1.045 | 1.045 | 5 |
| $\delta$ (RMS energy spread) | % | 10.5 | 20 | 20 | 16 | 16 | 35 |
| Luminosity | $\times 10^{34}$ | 4.9 | 7.9 | 7.9 | 6.1 | 6.1 | 5.9 |
| Photons/electron | | 1.95 | 2.1 | 2.1 | 1.2 | 1.2 | 2.2 |
| No. of coh. Pairs at IP | | 0 | $2\times 10^4$ | $2\times 10^4$ | $7.9\times 10^5$ | $7.9\times 10^5$ | $6.8\times 10^8$ |
| Incoh. pairs at IP | | 383 | 49 | 49 | 5 | 5 | $3\times 10^5$ |
| No. of cavities New + Existing | $\times 10^3$ | 11 + 16 | 27 + 11 | 21 +11 | 43 +0 | 37.5 +0 | 160 (0.25 m ea.) |
| No. of klystrons New + Existing | | 460 +360 = 820 | 820+460 =1280 | 755+425= 1180 | 690+820= 1500 | 1680+ 820 =2500 | |
| $Q_L$ (for new cav) | | $5.6\times 10^6$ | $8\times 10^6$ | $5.\times 10^6$ | $8\times 10^6$ | $1\times 10^7$ | |
| Input power (new cav) | kW | 350 | 365 | 460 | 300 | 550 | |
| New + Existing = Total Linac Length | km | 16+ 22 = 38 | 14 +38 = 52 | 6+38 =44 | 19+38 =57 | 12+38 = 50 | 42 |